 \documentclass[12pt,showpacs]{revtex4}
   
\IfFileExists{srcltx.sty}{\usepackage[active]{srcltx}}

\usepackage{graphicx}
\usepackage{epsfig}
\def\gtap{\mathrel{ \rlap{\raise 0.511ex \hbox{$>$}}{\lower 0.511ex
   \hbox{$\sim$}}}} \def\ltap{\mathrel{ \rlap{\raise 0.511ex
   \hbox{$<$}}{\lower 0.511ex \hbox{$\sim$}}}}
\newcommand{\beq}{\begin{equation}}

\newcommand{\eeq}{\end{equation}}
\newcommand{\bea}{\begin{eqnarray}}
\newcommand{\eea}{\end{eqnarray}}

\newcommand{\half}{\frac{1}{2} }

\newcommand{\eV}{\mbox{$ \ \mathrm{eV}$}}

\newcommand{\bz}{\bar{z}}
\newcommand{\sg}{\sigma}
\newcommand{\tw}{\tilde{\omega}}
\newcommand{\bmalpha}{\mbox{\boldmath $\alpha $}}
\begin{document}

\title{{\bf \Huge Dirac oscillator in an external magnetic field}}

\vspace{.3in}

\author{{\bf\large  Bhabani Prasad Mandal\footnote{Email: bhabani.mandal@gmail.com;
\ \ Phone No: 91-542-6701545 }}}
\author{{\bf \large Shweta Verma\footnote{Present Address: Raja Ramanna 
Centre for  Advanced Technology, Indore-452013}}}
\affiliation{Department of Physics,Banaras Hindu University,Varanasi-221005, India }


\vspace{1.2in}

\begin{abstract}

 We show that 2+1 dimensional Dirac oscillators   in an
external  magnetic field is mapped onto the  same  with reduced
 angular frequency in absence of  magnetic field. This  can be used to study the atomic
 transitions in a radiation field. Relativistic Landau levels
 are constructed explicitly. Several interesting  features of this
 system are discussed.
\end{abstract}

\pacs{ 31.30.J-; 03.65.Pm}


\maketitle
{\bf Keywords: Dirac Oscillator, Jaynes-Cumming Model, Relativistic Landau Levels,
Graphene}

\newpage

 Dirac oscillator \cite{do} was introduced for the first time   by Moshinsky and
Szczepaniak by adding the linear term $- imc\omega \beta \mbox{\boldmath $\alpha $} \cdot
{\bf r}$ to the usual Dirac Hamiltonian  for a free particle and 
 described by the Hamiltonian,
\beq H= c\mbox{\boldmath $\alpha$}\cdot ({\bf p}-im\omega\beta
{\bf r}) + \beta m c^2 \label{do}, \eeq
where {\boldmath $\alpha $},  $\beta $ are usual Dirac matrices,  $m$ is the
rest mass of the particle and $c$ is the speed of light.
 In the nonrelativistic limit, it reduces to
a quantum simple harmonic oscillator ( with  mass m, angular frequency, $\omega $)
 with a  strong spin-orbit coupling. The Dirac
oscillator has attracted a lot of attention and has found many physical applications in
various  branches of physics \cite{had,qcon,zit,pc,noc,sym1,mag,sym,ab,jella,ee,2d,new,new1,new2,new3,
new4}.
 Initially it was introduced in the context of
many body theory, mainly in connection with quark confinement models in
quantum chromodynamics (QCD) and in the context of relativistic quantum
mechanics \cite{qcon}. Later the  subsequent studies have revealed several exciting
properties connected to the symmetries of the theory \cite{sym1,sym}. Most exciting properties
of Dirac oscillator is its connection to quantum optics \cite{sym1,noc}.
 In 2+1 dimensions
it maps into Anti-Jaynes-Cummings (AJC)model  \cite{jc,ajc,rwa}  which describes the atomic
transitions in a two level system.

In this present article we have studied Dirac oscillator in presence of
 external magnetic field.
The dynamics of the Dirac oscillator in the  external uniform magnetic
 field, ${\bf B}$
is  governed by the Hamiltonian,

\beq
H= c{\bmalpha}\cdot ({\bf p}-\frac{e{\bf A}}{c} -im\omega\beta {\bf r}) + \beta m c^2
\label{dom}.
\eeq
${\bf A}$ is the vector potential which produces the external magnetic field
and $e$ is the charge of the Dirac oscillator which is considered to be an
 electron in this work.
 This theory of 2+1 dimensional Dirac oscillators in presence of
external transverse magnetic field is mapped onto the theory of Dirac
 oscillators without magnetic field in the same space-time dimension and described
by the Hamiltonian,
\beq
H^\prime= c{\bmalpha}\cdot ({\bf p} -im\tw\beta {\bf r}) + \beta m c^2
\label{dowm}.
\eeq
 The 
external magnetic field causes the angular frequency of the Dirac oscillator
to change  by  half of the cyclotron frequency. 
 We have solved 2+1 dimensional  Dirac equation corresponding to 
Hamiltonian  in  Eq. ( \ref{dom}) for Dirac oscillator
 in the presence of external, constant  magnetic field exactly and have obtained
 energy eigenvalues
and eigen functions analytically. Relativistic Landau levels \cite{lll} are
constructed explicitly. Infinitely  degenerate  lowest Landau level are constructed
in the coordinate basis in a very simple manner.
  Further, we derive exact mapping between this
relativistic model and AJC model so widely used in quantum
optics. This relativistic system has many peculiar properties some of which
 are similar to that of 
graphene \cite{grar,nat,gra2,gra1,gra}, monolayer of carbon atoms.

For the sake of simplicity, we consider the magnetic field, ${\bf B}$
along the $z$-direction.
The  vector potential for this particular magnetic field
can be chosen in the  symmetric gauge as
 ${\bf A}= ( -\frac{B}{2}y,
\frac{B}{2}x, 0). $ Further we are interested to study the system in two space
and one time dimension
 where  Dirac matrices can be chosen  in terms of
$2\times 2 $ Pauli spin  matrices. We consider one of the widely used choices 
of Dirac matrices as
    $\alpha_x =\sg_x , \alpha_y = \sg_y \ $ and $ \ \beta =\sg_z .$
 The Hamiltonian in Eq.(\ref{dom}) can be written nicely in terms of complex
 coordinate
$z=x+iy$ and conjugate momentum, $p_z$ , as
\beq
H=\left ( \begin{array}{c c} mc^2 & 2cp_z+im\tw c\bz \\ 2cp_{\bz}-im\tw
cz & -mc^2 \end{array} \right )
\label{ham},
\eeq

with
\bea
 p_z &=&-i\hbar\frac{d}{dz} = \frac{1}{2}(p_x-ip_y) \nonumber ,\\
 p_{\bz} &=-& i\hbar\frac{d}{d\bz} = \frac{1}{2}(p_x+ip_y),\ \mbox{  and }
\label{pz1} \\
\tw &=&\omega - \frac{\omega_c}{2}\ \ \ \mbox{ where }  \omega_c =\frac{|e|B}{mc}
\label{pz},
\eea

$\omega_c $ is the cyclotron frequency.
These  complex coordinates and their canonical conjugate momenta satisfy the
following commutation relations, i.e.
\beq
[z,p_z] = i\hbar = [\bz, p_{\bz}],
\ ;\ [\bz , p_z]   = 0=[ z ,p_{\bz}].
\label{pz2}
\eeq
We define the following creation and annihilation operators as
\bea
a&=& \frac{1}{
\sqrt{m\tw\hbar}}p_{\bz}-\frac{i}{2}\sqrt{\frac{m\tw}{\hbar}}z \nonumber ,
\\
a^\dagger &=& \frac{1}{
\sqrt{m\tw\hbar}}p_{z}+\frac{i}{2}\sqrt{\frac{m\tw}{\hbar}}\bz
\label{aa}.\eea It can be verified using Eqs. (\ref{pz2}) and (\ref{aa}) that
$[a,a^\dagger] =1; \ [a,a]=0=[a^\dagger ,a^\dagger ]$.

The Hamiltonian in Eq.(\ref{ham}) further is expressed in terms of creation and 
annihilation operators as  
\beq
H=\left ( \begin{array}{c c} mc^2 & 2c\sqrt{m\tw\hbar}~a^\dagger
\\ 2c\sqrt{m\tw\hbar}~a  & -mc^2 \end{array} \right )
\label{ham1}.
\eeq


The complete solution of the Dirac oscillator in the transverse
magnetic field is obtained by solving the time independent Dirac equation, $H\Psi 
=E\Psi $ corresponding to the above Hamiltonian. The relativistic Landau 
levels  are 
given as,
 \beq  E_n^{\pm} = \pm mc^2\sqrt{1+\frac{4\hbar\tw}{mc^2}n} ;\ \ \ \ n= 0,1,2 \cdots 
\label{en}
\eeq
with normalized negative and positive energy states are given by
\beq
|\Psi_n^{\pm}>\  =\  c_n^{\pm} |n; \ \half> + d_n^{\pm}|n-1;\  -\half>
\label{wf},
\eeq
with $c^{\pm}_n =\pm\sqrt{\frac{E_n^+\pm mc^2}{2E_n^+}}$ and
 $ d_n^\pm =\mp\sqrt{\frac{E_n^+\mp
mc^2}{2E_n^+}} $. We adopt the notation for the state as $|n, \half
m_s>\equiv  \psi_n(z,\bz)\chi_{m_s}$ where $n$ is the
 eigenvalue 
 for the number operator, $a^\dagger
a$ and  $m_s = \pm 1 $ are the eigenvalues of the operator $\sg_z$
i.e. $\sg_z|\half m_s> = m_s|\half m_s>$, and $\psi_n(z,\bz)$  is the space
part of the wave function in the coordinate representation whereas 
$\chi_{m_s}$ is the spin part of the wave function. These states indicate the
entanglement between orbital and spin degrees of freedom in the Dirac 
oscillator problem.

 The lowest Landau levels [LLL] are obtained by using the condition,
 \beq a \psi_0(z,\bz) =0.
\eeq Equivalently  in coordinate representation,the space part  wave function ($\psi_0(z,\bz)$) of the LLL
 satisfies
\beq
(\frac{\partial}{\partial\bz} +\frac{m\tw}{2\hbar}z) \psi_0(z,\bz) =0.
\label{lll}
\eeq
By substituting
\beq \psi(z,\bz) = e^{-\frac{m\tw}{2\hbar} z\bz} u_0(z,\bz)
\label{sub},
\eeq
 we further obtain
 \beq \frac{\partial}{\partial\bz}u_0(z,\bz)=0,
\label{lll1}
\eeq  as the defining rule for the LLL.
We obtain the LLL in the coordinate representation, which is infinitely degenerate, as
\beq
\psi_0 (z,\bar{z})=
 z^l e^{-\frac{m\tw}{2\hbar}z\bz},\ \ \ \ l=0,1,2\cdots ,
\label{lll2}
\eeq
 as
$u_0(z,\bz)$ is analytic function and monomials $z^l $ with $ l=0,1,2 \cdots $
 can  serve as a linearly independent basis. The first excited state and
other higher states in the coordinate space  can be obtained by applying
$a^\dagger $ given in Eq. (\ref{aa}) repeatedly.

We observe several interesting features  of this system and study their
implications and consequences. Firstly we observe that the 2+1 dimensional
 Dirac oscillator having angular frequency, $\omega $ in the transverse
magnetic field  maps into a 2+1 dimensional Dirac oscillator in absence
of magnetic field but having a different angular frequency, $\tw = \omega
-\half \omega_c$. The effect of transverse magnetic field in the system of
2+1 dimensional Dirac oscillator is to decrease the angular frequency by half
of the cyclotron frequency only and the entire dynamics  remains unchanged
except a special choice on the strength of the magnetic field, $B=\frac{2mc\omega}{|e|}$
for which $\tw =0$ and Dirac oscillator stops oscillating. This is highly non-trivial
situation and the dynamics is completely different from the dynamics of the system
without magnetic field.  

 To show this result we consider the Hamiltonian
for Dirac oscillator without magnetic field given in Eq. (\ref{do}) in 2+1 dimension and
 express it using
complex coordinates and their canonically conjugate momenta listed in Eq.
 (\ref{pz1}) as
 \beq
H=\left ( \begin{array}{c c} mc^2 & 2cp_z+im\omega c\bz \\ 2cp_{\bz}-im\omega
cz & -mc^2 \end{array} \right )
\label{bham}
\eeq
 This Hamiltonian is exactly same (except $\omega$  is replaced by $\tw $ ) as the Hamiltonian given in Eq. (\ref{ham}) which
  describes the  Dirac oscillator
 in
the transverse magnetic field in 2+1 dimensions.
 Dirac oscillator in presence of external fields is also considered  by others \cite{zit,mag,ab,ee} to discuss
different issues, in particular, to study Aharanov-Bohm effect \cite{ab}, effect of electric field \cite{ee}.
  2+1 dimensional Dirac oscillator in the transverse magnetic field is also considered
 in Ref. \cite{zit} to
   discuss different aspect
of chiral quantum phase transition in such a system.    

 Interestingly we note the spectrum of Landau levels in Eq. (\ref{en}) are
non-equidistant in energy and proportional to the $\sqrt{nB}$
for sufficiently large transverse  magnetic field. This feature is quite
different from the conventional two dimensional electronic system.
 However this feature is very similar to the surprising  material graphene,
 the sheet of carbon atoms \cite{grar,nat,gra2,gra1,gra}. The peculiar 
properties of graphene
mainly stem from its linear dispersion near Fermi surface, and the chiral
nature of electronic states entangling the momentum and pseudo-spin of the
quasi-particle arising due to sublattice structure of carbon atoms in graphene.
This entanglement of states in graphene is exactly similar to the entanglement of
 orbital and spin states  in Dirac oscillator given in Eq. (\ref{wf}). Relativistic
Landau Levels are experimentally realized \cite{new8}.  

Now we would like to show the connection of this model to the AJC model
so widely used in the study of quantum optics.The Hamiltonian corresponding to
the simple version of AJC model is given as,
\beq
H_{AJC}= g(\sg ^- a +\sg^+a^\dagger) + \sg_z mc^2 ,
\label{ajc}
\eeq
On the other hand the JC model is described by the Hamiltonian,
\beq
H_{JC}= g(\sg ^+ a +\sg^-a^\dagger) + \sg_z mc^2 ,
\label{jc}
\eeq
which is quite different from Hamiltonian of AJC model as the positions of
creation and annihilation operators are interchanged.
The Hamiltonian in the Eq. (\ref{ham1}) can be expressed exactly in the
form of $H_{AJC}$ as given in Eq. (\ref{ajc}) using Pauli spin
matrices 
where $\sg^{\pm}=\half (\sg_x\pm i\sg_y) $ are usual spin raising and lowering
 operators and
  $g= 2c\sqrt{m\tw\hbar}= 2c\sqrt{m\hbar(\omega-\frac{|e|B}{2mc})}$, coupling
 between spin and orbital degrees
of freedom which  depends on the strength of the magnetic field.
 It is very interesting to note how  two completely different
 theories are related. This connection is established for arbitrary strength of the magnetic field hence
differs with some earlier work \cite{zit}. This provides an alternative approach to study atomic
 transitions in
two level system using relativistic quantum mechanical model, in particular
using the theory of Dirac oscillators in the transverse magnetic fields.
 The connection between relativistic quantum mechanics and quantum optics
using trapped ions is realized experimentally in Ref. \cite{new9}

We further observe that the zitterbewegung frequency for the 2+1 dimensional Dirac
oscillator in the external magnetic field depends on the strength of the magnetic
field. To show this we start with some initial pure state at $t=0 $,
\beq
  |n,\half>(0)= \frac{E_n^+}{mc^2}[d_n^+|\Psi_n^->-d_n^-|\Psi_n^+>]
\label{ini}.
\eeq
This equation is obtained from the Eq.(\ref{wf}) by eliminating the
 state $|n-1;-\half>$.
The Eq. (\ref{ini}) shows that the  starting initial state is 
 a superposition of both  the positive
and negative energy
solutions. The time evolution of this state can be written as

\beq
  |n,\half>(t)= \frac{E_n^+}{mc^2}[d_n^+|\Psi_n^->e^{i\omega_nt}-d_n^-|\Psi_n^+>e^{-i\omega_n t}]
\label{tt},
\eeq
where $\omega_n $ is the frequency of oscillation between positive and negative energy
solutions and given as
\beq
\omega_n = \frac{E_n^+}{\hbar} = \frac{mc^2}{\hbar}\sqrt{1+\frac{4\hbar\tw}{mc^2}n} \mbox{ with }
\tw = \omega -\frac{|e|B}{2mc}\label{ww}
\eeq
Substituting $\Psi^+$ and $\Psi^- $ and the constants $d^+_n$ and $d^-_n$ in Eq. (\ref{tt}) we obtain,
\bea
|n,\half>(t)&=& [\cos{(\omega_n t)}-\frac{imc^2}{E_n^+}\sin{(\omega_nt)}]|n,\half>(0)\nonumber \\
&+& i\sqrt{\frac{{E_n^+}^2-m^2c^4}{m^2c^4}}\sin{(\omega_nt)}|n-1,-\half>(0)\nonumber \\
\eea
This equation shows the oscillatory behavior between the states $|n,\half >$ and 
$|n-1,-\half>$  which is exactly
similar to atomic Rabi oscillations \cite{rabi,rwa,ajc} occurring in the JC/AJC models. Rabi frequency is given by
$\omega_n$ given in Eq. (\ref{ww}).

We would like mention that in the limit magnetic field, $B\longrightarrow 0 $ all these results are
 consistent with results of Dirac oscillator ( with out
magnetic field) in Ref. \cite{noc} 

 In conclusion, we have studied  2+1 dimensional Dirac oscillator in the presence of  transverse external
 magnetic field by defining
suitable creation and annihilation operators in terms of  properly chosen canonical pairs of coordinates 
and momenta.We have  shown that the Dirac oscillator in  magnetic field is mapped
onto Dirac oscillator without magnetic field but with different angular frequency. The frequency changes by half the 
cyclotron frequency.
 The relativistic Landau  levels and corresponding eigenstates have been calculated
explicitly by solving corresponding Dirac equation. Lowest Landau levels have been constructed exactly 
in the coordinate  space representations which
shows the infinite degeneracy of such state in a simple manner. We have observed several interesting
 features of such system.
Some of the features are quite similar to those shown by graphene. 
In particular, the orbital and spin degrees of freedom  are entangled to each other, the Landau levels are
 non-equidistant
in energy and the level separations are proportional to $\sqrt{B}$ for large magnetic field.
  The connection between AJC model in quantum optics with Dirac oscillator in magnetic field is 
established in this work in a simple manner without taking any limit on the strength of the magnetic field.
 This will allow to study atomic transitions in quantum optics using the 
relativistic quantum mechanical models in presence of magnetic field.
The Zitterbewegung   frequency has been calculated for the Dirac oscillator in magnetic field and has shown to vary as
$\sqrt{nB}$ for sufficiently large magnetic field. This oscillation between the negative and positive 
energy states is similar to the Rabi oscillation in the two level systems. It will be interesting to study
graphene in Dirac oscilator potential in presence of transverse magnetic field.

\end{document}